\documentclass[preprint,longabstract]{aastex}

\shorttitle{CLASH z=11 candidate}
\shortauthors{Pirzkal et al.}

\begin{document}

\title{Not In Our Backyard: Spectroscopic Support for the CLASH z=11 Candidate MACS0647-JD}
\email{npirzkal@stsci.edu}

\author{Nor Pirzkal\altaffilmark{1},
Dan Coe\altaffilmark{1}, 
Brenda L. Frye\altaffilmark{2}, 
Gabriel Brammer\altaffilmark{1},
John Moustakas\altaffilmark{3},
Barry Rothberg\altaffilmark{4,5},
Thomas J. Broadhurst\altaffilmark{6}, 
Rychard Bouwens\altaffilmark{7}, 
Larry Bradley\altaffilmark{1}
Arjen van der Wel\altaffilmark{8}, 
Daniel D. Kelson\altaffilmark{9}, 
Megan Donahue\altaffilmark{10}, 
Adi Zitrin\altaffilmark{11,12},  
Leonidas Moustakas\altaffilmark{13}, 
Elizabeth Barker\altaffilmark{1}} 

\altaffiltext{1}{Space Telescope Science Institute, 3700 San Martin Dr., Baltimore, MD 21218, USA}
\altaffiltext{2}{Steward Observatory/Department of Astronomy, University of Arizona, 933 N Cherry Ave, Tucson, AZ 85721, USA}
\altaffiltext{3}{Department of Physics and Astronomy, Siena College, 515 Loudon Road, Loudonville, NY 12211}
\altaffiltext{4}{LBT Observatory, University of Arizona, 933 N.Cherry Ave,Tucson AZ 85721, USA}
\altaffiltext{5}{George Mason University, Department of Physics \& Astronomy, MS 3F3, 4400 University Drive, Fairfax, VA 22030, USA}
\altaffiltext{6}{Tel Aviv University - Wise Observatory}
\altaffiltext{7}{Universiteit Leiden}
\altaffiltext{8}{Max-Planck-Institut fur Astronomie, Heidelberg}
\altaffiltext{9}{Carnegie Institution of Washington}
\altaffiltext{10}{Dept of Physics and Astronomy, Michigan State University, East Lansing, MI 48824, USA}
\altaffiltext{11}{Hubble Fellow}
\altaffiltext{12}{Cahill Center for Astronomy and Astrophysics, California Institute of Technology, MC 249-17, Pasadena, CA 91125, USA}
\altaffiltext{13}{Jet Propulsion Laboratory, M/S 169-506, 4800 Oak Grove Drive, Pasadena, CA 91109}

\begin{abstract}
We report on our first set of spectroscopic Hubble Space Telescope observations of the $z\approx11$\ candidate galaxy strongly lensed by the MACSJ0647.7+7015 galaxy cluster. The three lensed images are faint and we show that these early slitless grism observations are of sufficient depth to investigate whether this high-redshift candidate, identified by its strong photometric break at $\approx 1.5 \mu m$, could possibly be an emission line galaxy at a much lower redshift. While such an interloper would imply the existence of a rather peculiar object, we show here that such strong emission lines would clearly have been detected. Comparing realistic, two-dimensional simulations to these new observations we would expect the necessary emission lines to be detected at $> 5 \sigma$\ while we see no evidence for such lines in the dispersed data of any of the three lensed images. We therefore exclude that this object could be a low redshift emission line interloper, which significantly increases the likelihood of this candidate being a bona fide $z\approx11$\ galaxy.
\end{abstract}


\section{Introduction}
The search for the first galaxies has been at the forefront of extragalactic astronomy ever since the initial discovery that QSOs are so distant that their UV spectral features were redshifted into the optical pass-bands \citep{Schmidt1963}.  Owing to their high space density, star forming  galaxies comprise a large fraction of the high-z galaxy population available to us at the highest redshifts.  
Gravitational lensing is an efficient method that allows the detection of high redshift star forming galaxy candidates that would otherwise have luminosities that are too faint to be detected using current space- or ground-based observatories \citep{Wyithe2011}. 

A strong $z\approx11$\ candidate was identified by \citet{Coe2013} in observations of the $z = 0.591$\ galaxy cluster MACSJ0647.7+7015. These observations were made as part of the Cluster Lensing And Supernova survey with Hubble \cite[CLASH,][]{Postman12} program. Analysis of three separate lensed images (JD1, JD2 and JD3) of the faint red object in near infrared pass-bands led \citet{Coe2013} to conclude that this target was a strong candidate for a $z\approx11$\ galaxy. 
Three images of the same high redshift candidate were identified with magnification factors estimated to be $\approx 8, 7,$\ and 2 for the JD1, JD2, and JD3 images, respectively. The three images are magnified to F160W AB magnitudes of $\approx 25.9, 26.1,$\ and 27.3, respectively. All three images show no detected flux in observations taken in 15 different HST filters over a wavelength range of $0.2\mbox{--}1.4\mu m$. However, all three were detected in the F140W ($> 6\sigma$\ at $1.4\mu m$) and F160W ($>12\sigma$\ at $1.6 \mu m$) filters. While \citet{Coe2013} ruled out that this source could be a low-redshift interloper by considering a variety of possible low-redshift objects and the strength of the lensing model, their conclusion relied on the detection of this object in only two infrared broad band filters, while this source remained undetected at the longer wavelengths probed by relatively low signal-to-noise Spitzer/IRAC observations at $3.6\mu m$\ and $4.5\mu m$. 
The authors pointed out that the discovery of such an object was consistent with expectations extrapolated from $z\approx8$\ \citep{Bradley12} but is in conflict with the significantly lower than expected number of $z>9$\ detected by \citet{Bouwens14} and  \citet{Oesch13}. Results from  \citet{Bradley12,Oesch13} were indicative of  a dramatic buildup in the number of galaxies and of the cosmic star formation rate density over a very short period of time ($<200$\ Myr) between $z\approx10$\ and 8. This was later repeatedly confirmed by subsequent observations as well as re-analyses of earlier data  \citep[e.g.][]{Oesch14a,Oesch14b,Ishigaki14,Zheng14,Bouwens14}.
There is currently only one other multiply imaged galaxy at $z\approx10$\ \citep{Zitrin14}.The only other $z>11$\ candidate to date is the $z\approx12$\ candidate UDF12-3954-6284 \citep{Ellis13}. This source has only been detected in a single band and it can therefore be more easily explained away as being a lower-redshift interloper with strong nebular emission \citep{Pirzkal2013}. However, UDF12-3954-6284 remains a possible high-redshift candidate as no emission lines have been conclusively detected in subsequent observations and analyses \citep{Bouwens13, Brammer13, Capak13}.

Confirming the nature of high-redshift candidates requires spectroscopic observations. If some of the very few $z>9$\ candidates already identified were to be shown to be  intermediate redshift sources, then the observed deficit of galaxies at $z\approx10$ would be even greater, requiring an even more rapid buildup of early galaxies. Unfortunately, spectroscopic confirmation of high-redshift candidates has lagged behind due to the faintness of the sources, the apparent lack of Ly-$\alpha$\ emission in high-redshift galaxies, and the bright infrared sky background in ground based observations  \citep{Shibuya12,Ono12}. The unlensed Hubble Ultra Deep Field \citep[HUDF][]{Beckwith06} $z > 9$ candidates are significantly fainter ($29.3 <$ F160W AB $< 29.7$), prohibiting spectroscopic confirmation with current telescopes and limiting the spectroscopic studies possible with future telescopes.
Fortunately, cluster lensing magnifies images of the MACS0647-JD candidate such that space-based low-resolution spectroscopy is feasible. 

While the possibility that this object is a low redshift emission line galaxy is in contradiction with current lens models, spectroscopically ruling out that this object is not a low redshift star forming galaxy is important because our treatment of the nebular emission might be simplistic and incomplete.
There is empirical evidence for the existence of faint moderate redshift galaxies with very strong emission lines. \citet{Straughn11}  identified several extreme emission line objects in  Early Release Science observations using the G102 and G141 WFC3 grisms. Their Figure 3, for example, shows the observed spectrum of a F098M 26.87 AB magnitude source (ID397) at z=1.76 and which has a very strong [OIII]/H$\beta$\ doublet (${\rm 9.69\pm1.3\ 10^{-17} erg/s/cm^2}$, observed EW$\approx5000\AA$). This emission line is sufficient to produce a photometric break of nearly two AB magnitudes and is nearly as extreme as what would be required to reproduce the flux measured in  MACS0647-JD (${\rm \approx 5 \times 10^{-17} erg/s/cm^2}$).
See also \citet{Frye2012} for other examples of extreme emission line galaxies with large single
emission line fluxes of up to $1.38 \times 10^{-16}$erg s$^{-1}$ cm$^{-2}$.

Redshift confirmation is therefore essential to definitively rule out a low redshift emission line interloper (or perhaps discover the true nature of a very interesting galaxy).

\section{Observations}
This paper is based on a set of new Hubble Space Telescope near-infrared stilless spectroscopic observations of the MACSJ0647.7+7015 cluster. As part of a follow-up project, we will be observing this cluster for 12 orbits using the G141 grism on the Wide Field Camera 3 (WFC3). These observations are split in three epochs, revisiting this field at different position angle on the sky. The position angles were chosen to minimize the amount of spectral contamination in the spectra of our target and to provide stronger, independent detections of any spectral break or emission lines in the spectra of this source. We have obtained our first set of observations, which we present here. 
The  data were processed using the aXe data reduction package \citep{Pirzkal01,Kummel09}, using a custom sky subtraction step which includes different components for Zodiacal light, Earth limb, and stray light. More details on this sky subtraction can be found in \citet{Brammer14}. We used archival data to create mosaics of the field in the F105, F125W, F140W and F160W filters and to model the SED of every object in the field. This allowed us to model the spectral contamination caused by nearby objects using aXe's FluxCube quantitative estimates. This approach also allowed us to perform both regular and optimally weighted spectral extractions. Estimating spectral contamination is crucial when analyzing slitless spectroscopic  because overlapping spectra can cause features that look very similar to spectral breaks and zeroth dispersion orders of nearby sources  can also look similar to strong emission lines. A complete understanding of the field is therefore needed to rule out spectral contamination as the cause of any feature we detect in a spectrum.

\section{High redshift or Low Redshift Emission Line Interloper?}\label{SED}
While   \citet{Coe2013} suggest the most plausible explanation for the detection of MACS0647-JD in the
F140W and F160W filters is that the object is a lensed galaxy at z=11, it is not the only possible scenario.  Amongst the possible scenarios considered by the authors, one consists of the coincidental presence of two emission lines, one in each filter, to mimic a photometric break.

\begin{figure}
\centering
\includegraphics[width=5.5in]{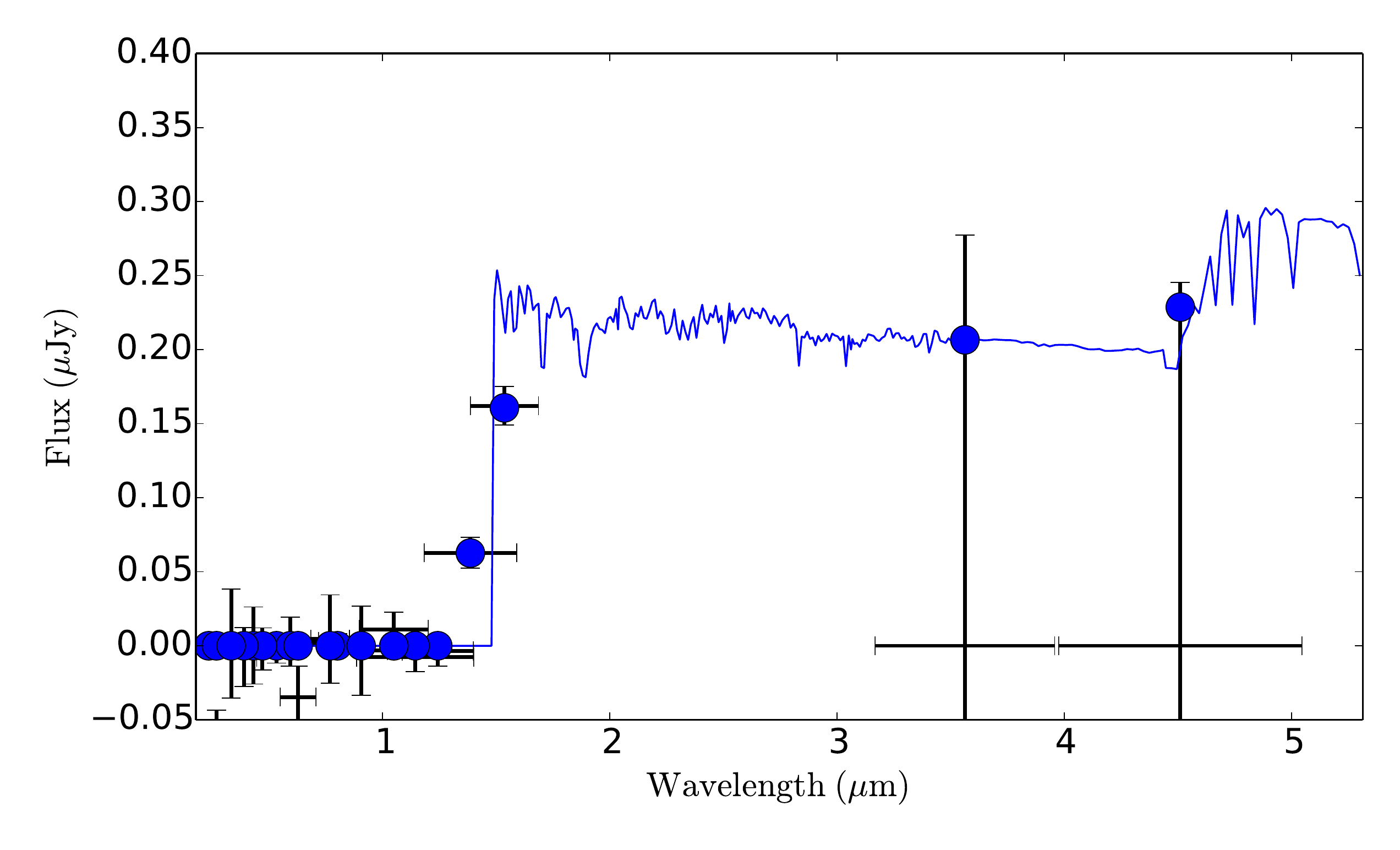} 
\caption{The observed broad band flux levels for JD1 (error bars) and the spectrum of a $z=11.2$, $10^{10} M_\sun$\ Lyman Break Galaxy. The model fluxes (blue circles) are in good agreement with the observations.}
\label{fit:z11}
\end{figure}

To further explore this, we started by re-fitting the data from Table 2 of \citet{Coe2013} using the Monte Carlo Markov Chain approach (MCMC)  described extensively in \citet{Pirzkal2012} and used to analyze other high-redshift candidates \citep{Pirzkal2013}.  Our approach uses \citet{Bruzual03} stellar population models, a Salpeter initial mass function, and SSP models. The free parameters are the redshift, stellar mass, stellar age, extinction level \citep{Calzetti00}, metallically and the ionizing photon escape fraction. This new analysis confirmed that the \citet{Coe2013} photometry were  better fit, especially at  bluer wavelengths, by a template consistent with a high-redshift Lyman break object at $z\approx11$. Figure \ref{fit:z11} demonstrates that  a $10^{10} M_\sun$\ galaxy at a redshift of $z\approx11$\ provides a very good fit. Although, as we show, such a model would require a significant amount of flux, about the same level that was detected in the F160W filter, in the IRAC bands at ${\rm 3.6\ and\ 4.5 \mu m}$, the IRAC data are not very sensitive and are consistent with a high-redshift Lyman-Break galaxy model.  

\begin{figure}
\centering
\includegraphics[width=5.5in]{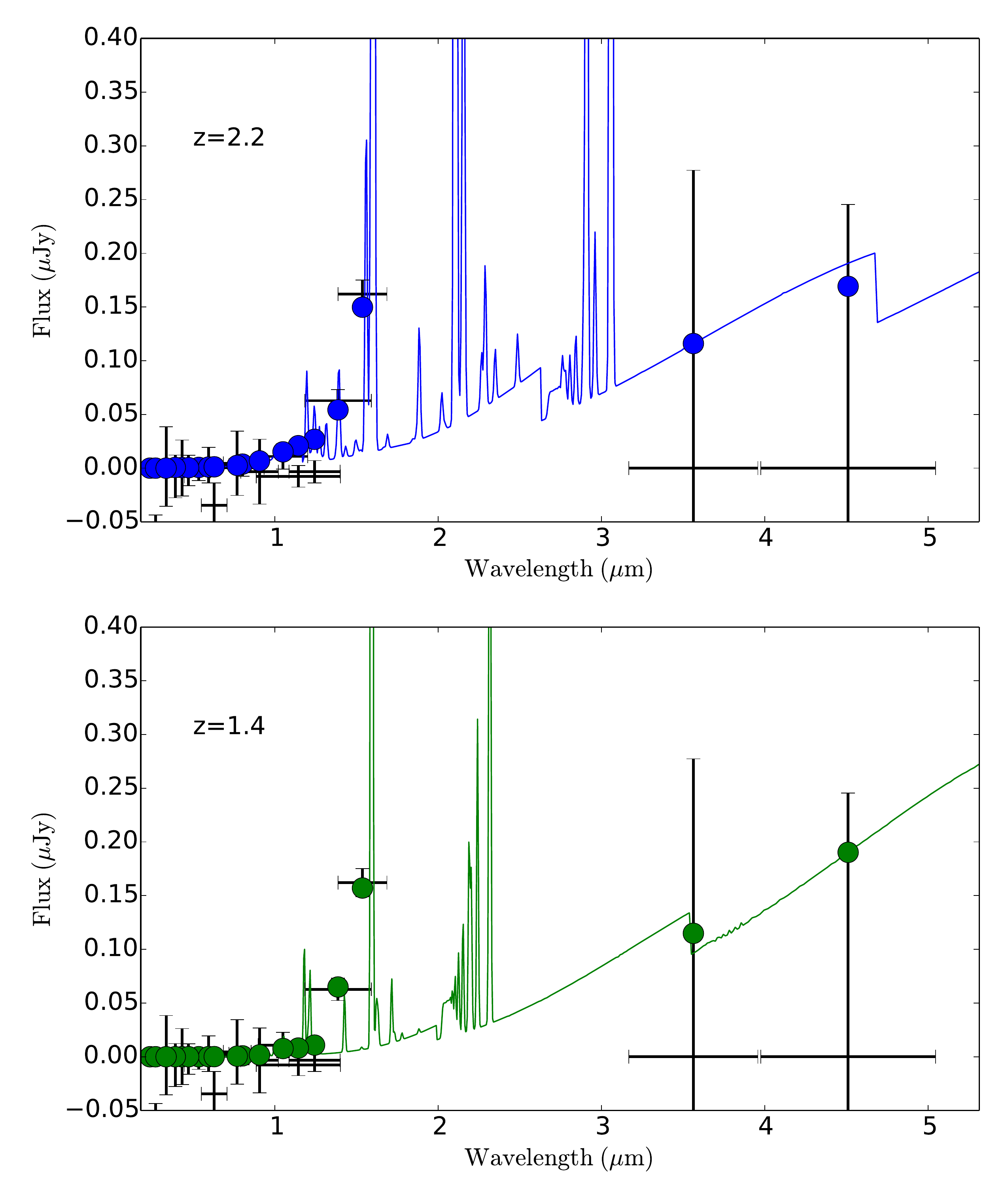} 
\caption{The observed broad band flux levels for JD1 (error bars) and the spectra of two young emission line objects at z=2.2 (Top Panel) and z=1.4 (Bottom Panel). The model fluxes (solid circles) are in good agreement with the observations, although the observed break is sharper than what the models allow when including the nebular continuum emission, as we did here. Models that reproduce the observed break can be generated by allowing a strong $z\approx1.4$\ H$\alpha$\ or a strong $z\approx2.2$\ [OIII] emission line to be the dominant source of the light detected using the HST F140W and F160W filters. Several such models at $z\approx1.4 \pm 0.05$\ and $z\approx2.2 \pm 0.05$ are in principle possible.\label{fit:emm}}
\end{figure}

Since the MCMC SED fitting approach allows us to fully explore our input parameter space, we can examine what low redshift models ($z<2.5$) best fit the photometric observations. However, this approach is ultimately limited by our handling of the nebular emission process, which we assume here to be the source of the high Equivalent Width (EW) lines in the spectra of our simulated star forming galaxies. In Figure \ref{fit:emm}, we show two distinct cases, one at $z\approx 1.4$\ and one at $z\approx 2.2$, where a young 
(few $10^6$\ year old)  star forming galaxy with strong nebular activity can reproduce the broad band observations. In our treatment of nebular emission, we have assumed the same ionizing photon escape fraction as for the nebular emission. The detailed description of how nebular emission is simulated can be found in \citet{Pirzkal2012}. Ultimately, it is the ratio of the nebular line emission to the continuum emission which limits how much of the
broad-band photometric break can be reproduced by the nebular emission lines of our models .  We also must resort to an amount of extinction large enough (${\rm A_v \approx 3\mbox{--}4}$) to quench the amount of blue light, including the nebular continuum light, so that such an object can remain undetected in broad band filters blue-ward below $\approx 1.5\mu m$. As Figure \ref{fit:emm} shows, a $z\approx 2.2$\ galaxy with H$\beta$\ + [O III] lines of combined flux of ${\rm \approx 5\times 10^{-17} erg/s/cm^2}$, with a rest-frame Equivalent Width (EW) of $\approx 3000\AA$, reproduces the observed fluxes, with [O III] $5007\AA$\ redshifted just red-ward of F140W, contributing the extra observed flux to F160W.
Similarly, a $z\approx1.4$\ galaxy with equally strong H$\alpha$\ emission could also possibly reproduce the observations.

Our multiple nebular emission lines model strongly constrains the redshifts at which we would expect 
a low redshift interloper to be $z\approx1.4 \pm 0.05$ (strong H$\alpha$\ emission) and $z\approx2.2 \pm 0.05$ (strong H$\beta$\ + [O III] emission).  
As shown in Figure 2, in order to produce the observe F140W flux levels, fainter emission lines are required.  
This places strong constraints on the wavelength range at which strong emission lines are expected.
Even if our detailed treatment of nebular emission was inaccurate, the observed photometric break implies 
the existence of some sort of bright emission line between $1.2 \mu m$\ to $1.7 \mu m$ (the combined wavelength 
coverage of the F140W and F160W filters). This is well within the bandpass of the G141 observations.

As demonstrated here, it is possible to create a scenario in which emission lines of a relatively low-z galaxy can mimic the photometric break of a higher redshift object. While these models would have to be cherry-picked and might appear to be unphysical and contrived, the simplest and most direct way to rule out that any of these scenarios are taking place is by demonstrating the absence of  strong emission lines in spectroscopic observations of this source.

\section{Slitless Observations}
The confirmation of a low-redshift emission line interloper would imply the existence of a rather peculiar galaxy. The HST WFC3 grism mode, and its  resolution of a few tens of angstroms per resolution element  is extremely adept at detecting emission line objects \citep[][]{Straughn11} and provides us with an opportunity to test the scenario proposed in the previous section.

\begin{figure}
\centering
\includegraphics[width=5.5in]{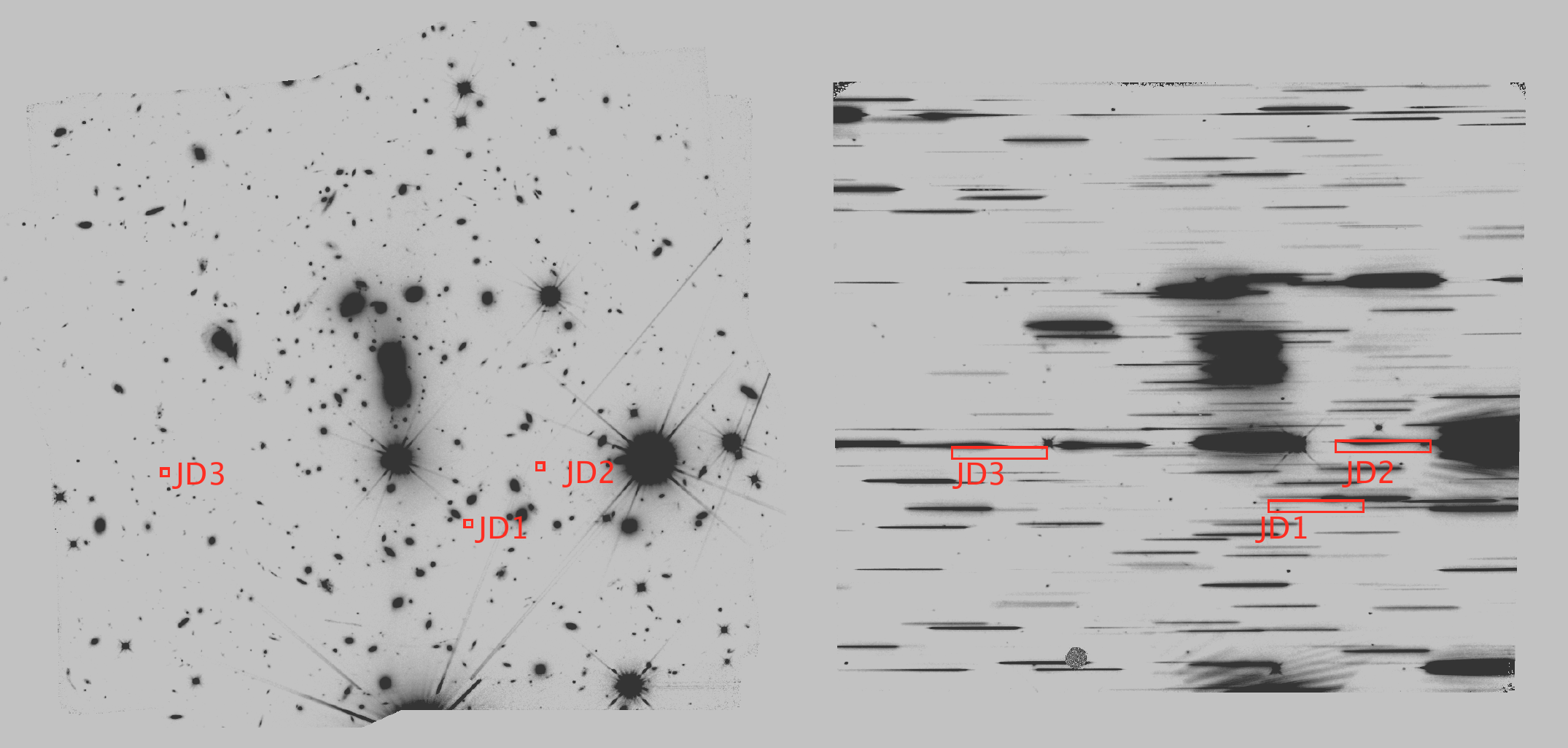} 
\caption{Left Panel: Composite (F105W+F125W+F140W+F160W) image of CLASH MACSJ0647. The three lensed images of MACSJ0647-JD are shown. Right Panel: The same field, dispersed using the HST G141 grism. The locations of the JD1, JD2 and JD3 lensed images are shown in the left Panel. The locations of the spectra of JD1, JD2 and JD3 in the G141 data are shown in the right Panel.}
\label{fig:MACS}
\end{figure}

As shown in the middle panel of Figure \ref{fig:MACS}, the WFC3 Grism field is extremely crowded and slitless spectroscopic observations are therefore difficult. Most spectra, especially the ones from fainter sources such as the three lensed images of MACS0647-JD, are contaminated by the spectra of nearby objects.
On the other hand, the WFC3 G141 grism is well understood, well calibrated, and WFC3 has has been remarkably stable over the past six years. We are therefore able to generate very accurate two-dimensional simulations of what this field would look when observed with the WFC3 G141 grism. The inputs of these simulations are mosaics of the entire MACSJ0647.7+7015  field using data from the CLASH program. Mosaics in the F105W, F125W, F140W and F160W can then be used to detect the footprint and estimate the spectral energy distribution of every object in the field. Simulations are then generated  by using the aXeSIM package, along with the photometric data and the latest WFC3 G141 calibration products. With the exception of the still poorly calibrated zeroth spectral order, the resulting simulations are a very good match to our observations. These simulations are similar to what is done when computing contamination and optimal extraction weigths as part of routine grism extraction \citep[e.g.][]{Pirzkal04, Brammer12}. The observations and the simulations are shown in the middle and right panels of Figure \ref{fig:MACS}, respectively.

\begin{figure}
\centering
\includegraphics[width=5.5in]{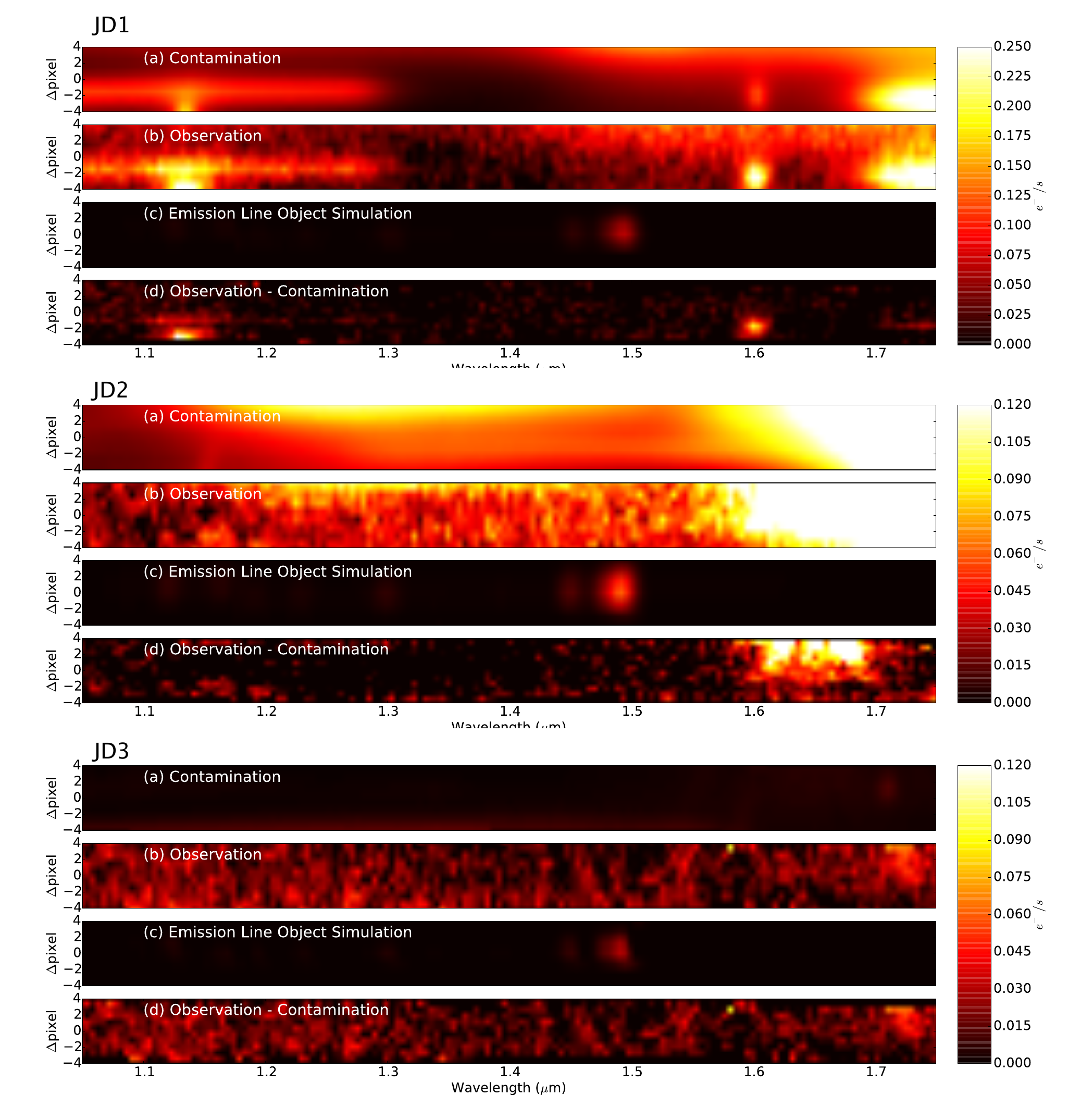} 
\caption{G141 observation of the lensed images JD1, JD2 and JD3. Panels (a) show the model of the spectral contamination from all of the sources in the field. Panels (b) show our observations. Panels (c) show our G141 simulations of what the lensed images would look like dispersed if its spectrum was that of the emission line object shown in Figure \ref{fit:emm}. As these simulations of a $z\approx2.2$ H$\beta$+[OIII] emission line galaxy show, an emission line at $\approx 1.5\mu m$ bright enough to reproduce the observed photometric break would be easily detected in our observations of JD1 and JD2. We see no evidence of such a strong emission line in Panels (d) where we show our contamination subtracted observations of JD1, JD2 and JD3. The contamination estimate is accurate and subtracts cleanly from the actual observations. The main residual are the two point like zeroth orders at 1.13$\mu m$\ and 1.6$\mu m$. This is expected as zeroth orders are relatively poorly calibrated and hence only approximately simulated. These are however located far away, particularly in the y-direction, from where we would potentially expect an emission line from JD1.}
\label{fig:JD123}
\end{figure}

Figure \ref{fig:JD123} qualitatively compares our shallow observations to our simulated dispersed image of the strong emission line source already shown in Figure \ref{fit:emm}. In these simulations, and since these three lensed images are partially resolved, we have conservatively set the size of the emission line region to be equal to the measured sizes of JD1, JD2 and JD3 in the broad band images. These sizes were measured from the available broad band imaging and identical to the ones listed in \cite{Coe2013}. Our simulated emission lines are therefore more spatially diluted than they would be if we had simply assumed an unresolved point source.
It is immediately obvious that any emission line(s) bright enough to reproduce the observed broad band photometric break would be detected in 3 orbit grism observations. As Figure \ref{fig:JD123} shows, there is no sign  of such an emission line in our slitless observations for either JD1, JD2 or JD3, demonstrating with little doubt that the CLASH $z\approx11$\ candidate is devoid of any bright emission lines in the wavelength range covered by the G141 grism. This significantly strengthens the case that this source is a $z\approx11$\ Lyman-break galaxy.

\begin{figure}
\centering
\includegraphics[width=5.5in]{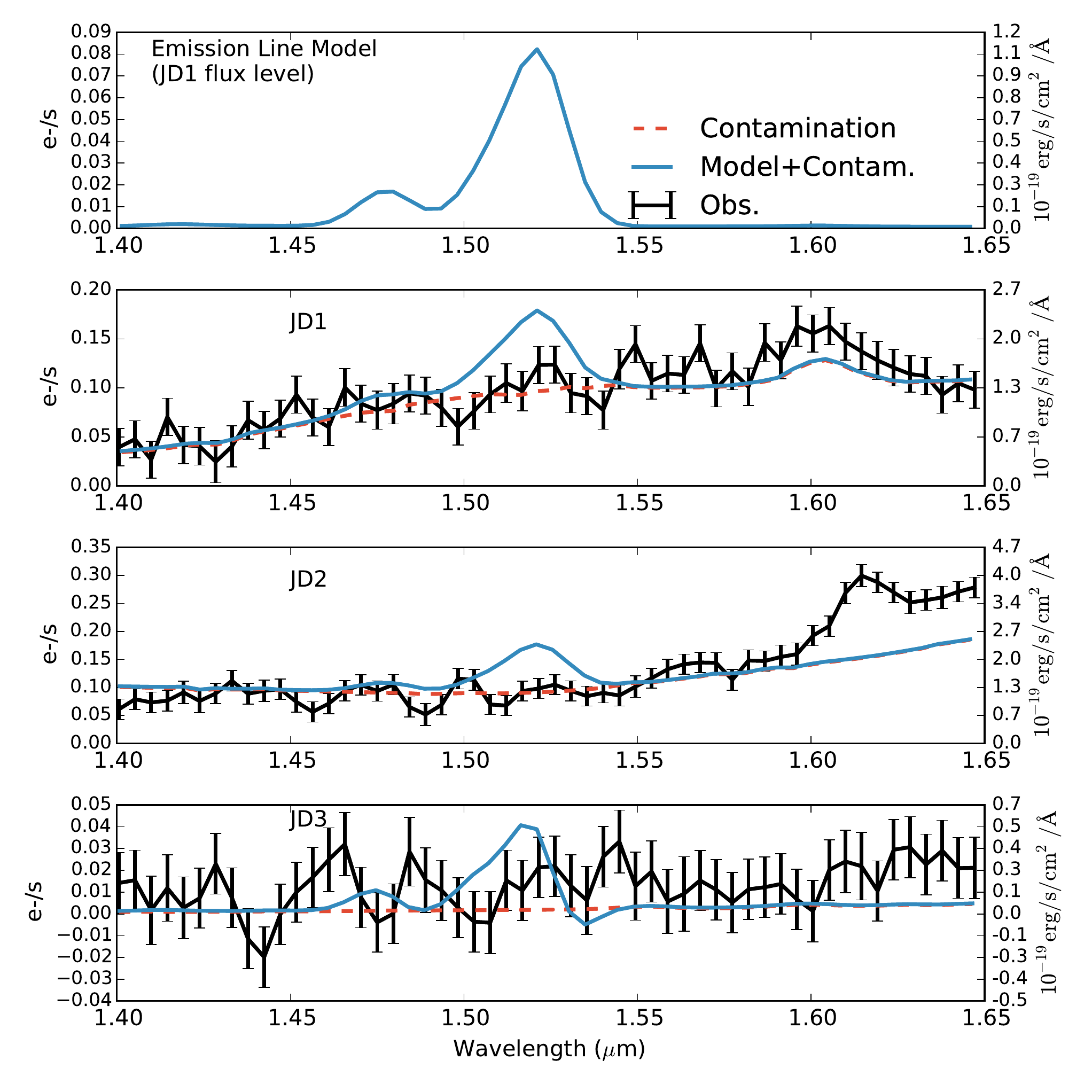} 
\caption{Extracted spectra of JD1, JD2 and JD3 using a narrow extraction window of $\approx1.5$\ pixel to maximize the signal-to-noise (line with error bars). We also show the emission line model spectrum (solid line) of a $z\approx2.2$ H$\beta$+[OIII]  line interloper in the top panel for JD1. This is the same spectrum shown in Figure \ref{fit:emm}. The expected contamination level is shown using a dash line. The contamination is caused by the overlapping spectra of other sources in the field. The emission line models for JD2 and JD3 are similar to the one shown in the top panel for JD1 when accounting for the slight difference in spatial size and lower fluxes.}
\label{spc:emm}
\end{figure}

For a more quantitative comparison, we show the extracted one-dimensional spectra of the three MACSJ0647-JD sources in Figure \ref{spc:emm}. As this Figure shows, the low spectral resolution of the G141 grism would make the emission lines shown in Figure \ref{fit:emm} less prominent and broader. These lines would nevertheless remain readily detectable at the $>5 \sigma$ level. As we described in Section \ref{SED}, there are actually narrow ranges of redshifts ($\Delta z \approx 0.05$) over which our models can reproduce the broad-band photometric break. While this would result in the emission line in Figure \ref{fig:JD123} to be shifted, it would remain easily detectable in at least JD1 and JD3, where contamination is either at longer wavelength, or vertically offset from where we would expect an emission line to be located.

\begin{figure}
\centering
\includegraphics[width=5.5in]{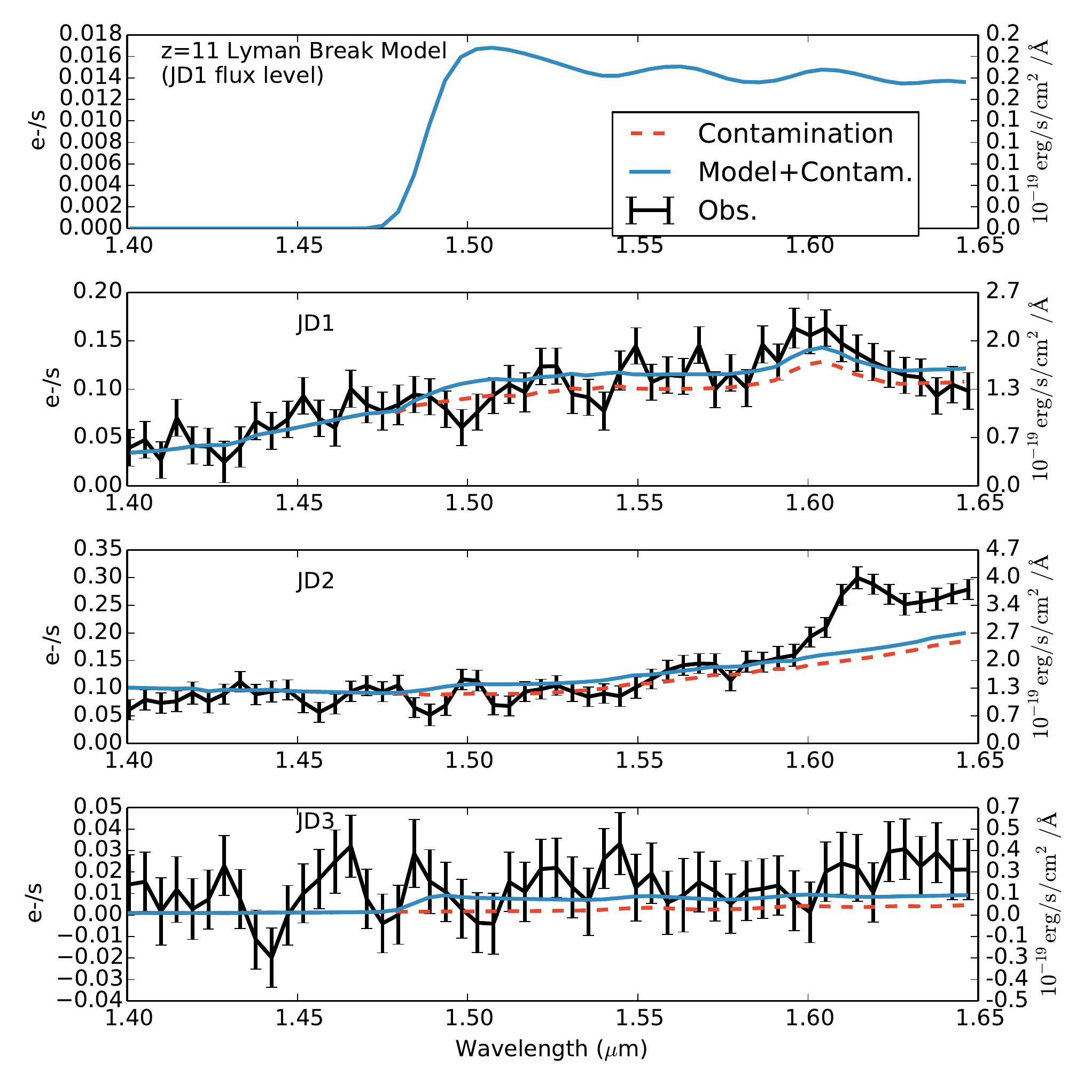} 
\caption{Extracted spectra of JD1, JD2 and JD3 using a narrow extraction window of $\approx1.5$\ pixel to maximize the signal-to-noise (line with error bars). We also show the emission line model spectrum (solid line) of a high-redshfit LBG in the top panel for JD1. This is the same spectrum shown in Figure \ref{fit:z11}. The expected contamination level is shown using a dash line. }
\label{spc:lbg}
\end{figure}

Figure \ref{spc:emm} shows the spectra in units of $e^-/s$ with the corresponding flux limits shown on the right axis. We show these extracted spectra in these units as their noise level is sky dominated and approximately constant ($\approx 0.02e^-/s$). Figures \ref{spc:lbg} shows the extracted spectra of the three lensed images as well as that of the LBG presented in Figure \ref{fit:z11}. The expected flux levels for JD1, shown in the top panel of this Figure are $\approx0.02e^-/s$ and the relatively high level of contamination (red dashed line) renders the detection of the continuum impossible in these observations. The forthcoming observations, chosen to avoid contamination,  will triple  the integration time and greatly improve our odds of detecting the continuum emission from this source.

\section{Conclusion}
We have presented results based on our shallow, first epoch slitless spectroscopic observations of the galaxy cluster MACSJ0647.7+7015. These data have allowed us to obtain the first spectroscopic observations of the three lensed images of the $z\approx11$\ LBG candidate MACSJ0647-JD of \cite{Coe2013}. Following a detailed modeling of the entire field of view, we showed that we can exclude a scenario in which the photometric break at $1.5\mu m$\ in the SED of this object could be caused by a large EW  emission line. Any emission line strong enough to produce the observed broad band photometric break would have been easily detected  in these observations. We conclude that this object remains an extremely strong candidate for a genuine $z\approx11$\ galaxy. Based on the simulations in this paper, we expect the continuum level red-ward of the Lyman Break to be detected in planned deeper future observations.

\acknowledgments
We would like to thank the referee for his or her comments and suggestions to improve the quality of this paper.
Support for AZ was provided by NASA through Hubble Fellowship grant \#HST-HF2-51334.001-A awarded by STScI. This work was supported in part by grant HST-GO-13317.13 from the Space Telescope Science Institute, which is operated by AURA under NASA contract NAS5-26555.

{}

\begin{thebibliography}{}
\bibitem[Beckwith et al.(2006)]{Beckwith06} Beckwith, S.~V.~W., Stiavelli, M., Koekemoer, A.~M., et al.\ 2006, \aj, 132, 1729
\bibitem[Bouwens et al.(2013)]{Bouwens13} Bouwens, R.~J., Oesch, P.~A., Illingworth, G.~D., et al.\ 2013, \apjl, 765, L16 
\bibitem[Bouwens et al.(2014)]{Bouwens14} Bouwens, R.~J., Bradley, L., Zitrin, A., et al.\ 2014, \apj, 795, 126 
\bibitem[Brammer et al.(2012)]{Brammer12} Brammer, G. B., van Dokkum, P. G., Franx, M., et al.\ 2012, ApJS, 200, 13
\bibitem[Brammer et al.(2013)]{Brammer13} Brammer, G.~B., van Dokkum, P.~G., Illingworth, G.~D., et al.\ 2013, \apjl, 765, L2
\bibitem[Brammer et al.(2014)]{Brammer14} Brammer, G.~B et al.. 2014, HST WFC3-ISR, TBD
\bibitem[Bradley et al.(2012)]{Bradley12} Bradley, L.~D., Trenti, M., Oesch, P.~A., et al.\ 2012, \apj, 760, 108 
\bibitem[Bruzual \& Charlot (2003)]{Bruzual03} Bruzual, G.,\& Charlot, S. 2003, MNRAS, 344, 1000
\bibitem[Calzetti et al. (2000)]{Calzetti00} Calzetti, D., Armus, L., Bohlin, R. C., et al. 2000, ApJ, 533, 682
\bibitem[Capak et al.(2013)]{Capak13} Capak, P., Faisst, A., Vieira, J.~D., et al.\ 2013, \apjl, 773, L14 
\bibitem[Coe et al.(2013)]{Coe2013} Coe, D., Zitrin, A., Carrasco, M., et al.\ 2013, \apj, 762, 32 
\bibitem[Ellis et al.(2013)]{Ellis13} Ellis, R.~S., McLure, R.~J., Dunlop, J.~S., et al.\ 2013, \apjl, 763, L7
\bibitem[Frye et al.(2012)]{Frye2012} Frye, B.~L., Hurley, M., Bowen, D.~V., et al.\ 2012, \apj, 754, 17 
\bibitem[Ishigaki et al.(2014)]{Ishigaki14} Ishigaki, M., Kawamata, R., Ouchi, M., et al.\ 2014, arXiv:1408.6903 
\bibitem[K\"{u}mmel et al.(2009)]{Kummel09}K\"{u}mmel, M., Walsh, J. R., Pirzkal, N., et al. 2009, PASP, 121, 59
\bibitem[Ono et al.(2012)]{Ono12} Ono, Y., Ouchi, M., Mobasher, B., et al.\ 2012, \apj, 744, 83
\bibitem[Oesch et al.(2013)]{Oesch13} Oesch, P.~A., Bouwens, R.~J., Illingworth, G.~D., et al.\ 2013, \apj, 773, 75
\bibitem[Oesch et al.(2014a)]{Oesch14a} Oesch, P.~A., Bouwens, R.~J., Illingworth, G.~D., et al.\ 2014, \apj, 786, 108 
\bibitem[Oesch et al.(2014b)]{Oesch14b} Oesch, P.~A., Bouwens, R.~J., Illingworth, G.~D., et al.\ 2014, arXiv:1409.1228 
\bibitem[Maraston (2005)]{Maraston05} Maraston, C. 2005, MNRAS, 362, 799
\bibitem[Pirzkal et al.(2001)]{Pirzkal01} Pirzkal, N., Pasquali, A., \& Demleitner, M. 2001, ST-ECF Newslett., 29, 5
\bibitem[Pirzkal et al.(2004)]{Pirzkal04} Pirzkal, N., Xu, C., Malhotra, S., et al.\ 2004, \apjs, 154, 501 
\bibitem[Pirzkal et al.(2012)]{Pirzkal2012} Pirzkal, N., Rothberg, B., Nilsson, K.~K., et al.\ 2012, \apj, 748, 122 
\bibitem[Pirzkal et al.(2013)]{Pirzkal2013} Pirzkal, N., Rothberg, B., Ryan, R., et al.\ 2013, \apj, 775, 11 
\bibitem[Postman et al.(2012)]{Postman12} Postman, M., Coe, D., Ben{\'{\i}}tez, N., et al.\ 2012, \apjs, 199, 25
\bibitem[Schmidt (1963)]{Schmidt1963} Schmidt, M. 1983, Nature, 197, 1040
\bibitem[Shibuya et al. (2012)]{Shibuya12} Shibuya, T., Kashikawa, N., Ota, K., et al.\ 2012, \apj, 752, 114 
\bibitem[Straughn et al.(2011)]{Straughn11} Straughn, A.~N., Kuntschner, H., K{\"u}mmel, M., et al.\ 2011, \aj, 141, 14 
\bibitem[Wyithe et al.(2011)]{Wyithe2011} Wyithe, J.~S.~B., Yan, H., Windhorst, R.~A., \& Mao, S.\ 2011, \nat, 469, 181 
\bibitem[Zheng et al.(2014)]{Zheng14} Zheng, W., Shu, X., Moustakas, J., et al.\ 2014, \apj, 795, 93 
\bibitem[Zitrin et al.(2014)]{Zitrin14} Zitrin, A., Zheng, W., Broadhurst, T., et al.\ 2014, \apjl, 793, LL12 
\end{thebibliography}
\end{document}